\documentclass[aps,pre,twocolumn,superscriptaddress]{revtex4}

\newif\ifpdf
\ifx\pdfoutput\undefined
\pdffalse 
\else
\pdfoutput=1 
\pdftrue
\fi

\ifpdf
\usepackage[pdftex]{graphicx}
\else
\usepackage{graphicx}
\fi
\usepackage{hyperref}

\begin{document}

\ifpdf
\DeclareGraphicsExtensions{.pdf, .jpg}
\else
\DeclareGraphicsExtensions{.eps, .jpg}
\fi

\def\hslash{\hbar}
\def\imag{i}
\def\grad{\vec{\nabla}}
\def\div{\vec{\nabla}\cdot}
\def\curl{\vec{\nabla}\times}
\def\DDt{\frac{d}{dt}}
\def\ddt{\frac{\partial}{\partial t}}
\def\ddx{\frac{\partial}{\partial x}}
\def\ddy{\frac{\partial}{\partial y}}
\def\lap{\nabla^{2}}
\def\divv{\vec{\nabla}\cdot\vec{v}}
\def\gradS{\vec{\nabla}S}
\def\vvec{\vec{v}}
\def\wc{\omega_{c}}
\def\<{\langle}
\def\>{\rangle}
\def\Tr{{\rm Tr}}
\def\Csch{{\rm csch}}
\def\Coth{{\rm coth}}
\def\Tanh{{\rm tanh}}
\def\g2{g^{(2)}}
\newcommand{\al}{\alpha}

\newcommand{\la}{\lambda}
\newcommand{\del}{\delta}
\newcommand{\om}{\omega}
\newcommand{\ep}{\epsilon}
\newcommand{\pd}{\partial}
\newcommand{\bra}{\langle}
\newcommand{\ket}{\rangle}
\newcommand{\bbra}{\langle \langle}
\newcommand{\kket}{\rangle \rangle}
\newcommand{\non}{\nonumber}
\newcommand{\be}{\begin{equation}}
\newcommand{\ee}{\end{equation}}

\title{Time-convolutionless master equation for mesoscopic electron-phonon systems}

\author{Andrey Pereverzev}
\email{aperever@mail.uh.edu}
\affiliation{Department of Chemistry, Center for Materials Chemistry, 
and the Texas Center for Superconductivity,
University of Houston \\ Houston, TX 77204}
\author{Eric R. Bittner}
\email{bittner@uh.edu}
\affiliation{Department of Chemistry, Center for Materials Chemistry, 
and the Texas Center for Superconductivity,
University of Houston \\ Houston, TX 77204}
\date{\today}

\begin{abstract}
The time-convolutionless master equation for the electronic
populations is derived for a generic electron-phonon Hamiltonian.  The
equation can be used in the regimes where the golden rule approach is
not applicable. The equation is applied to study the electronic
relaxation in several models with the finite number normal modes. For
such mesoscopic systems the relaxation behavior differs substantially
from the simple exponential relaxation.  In particular, the equation
shows the appearance of the recurrence phenomena on a time-scale determined by 
the slowest mode of the system.  The formal results are quite general and can be used 
for a wide range of physical systems. 
Numerical results are presented 
for a two level system coupled to a Ohmic and super-Ohmic baths, as well as for a
a model of charge-transfer dynamics between semiconducting organic polymers. 
\end{abstract}


\maketitle
\section{Introduction}
%
The Pauli master equation and the Redfield equation have long been
applied to the study of relaxation dynamics in electronic systems
\cite{Redfield, May, Breuer}. These equation were originally derived
using semi-heuristic arguments.  Hwwever, with the advent of
projection operator techniques in 1960's \cite{Nakajima, Zwanzig,
Prigogine} it became clear that such master equations could be
obtained from formally exact generalized master equations through a
series of well defined approximations.  Such approximations are
usually justified when the system of interest is coupled to a
thermostat with a large number of degrees of freedom.  In this limit,
the resultant transition probabilities in the Pauli master equation
are identical to those obtained by the less rigorous golden rule
approach \cite{May}.

The case when the thermostat is mesoscopic (i.e. with a finite number
of modes) is less well studied.  In general, for this case the time
independent transition probabilities of the golden rule type may not
be well defined since the kernels of the generalized master equations
do not vanish in the limit $t\to\infty$.  Our goal in this paper is to
develop a the convolutionless generalized master equation suitable for
a mesoscopic electron-phonon system using the fact that this equation
becomes formally exact for arbitrary system sizes.  In contrast to the
derivation of the Pauli master equation, the only approximation that
we will use is that of weak coupling.  As a result we will arrive at a
master equation with time dependent rate coefficients.
  
\section{Hamiltonian and the initial state}
A wide class of the electron-phonon systems can be described by the
following Hamiltonian ($\hbar=1$) 
\begin{eqnarray} H=\sum_n\epsilon_n |n\ket\bra
n|+\sum_{nmi} g_{nmi}|n\ket\bra m|(a^{\dagger}_i+a_i)
+\sum_i\om_ia^{\dagger}_ia_i. \label{Ham} 
\end{eqnarray} 
Here $|n\ket$'s denote
electronic states with vertical energies $\epsilon_n$, $a_i^{\dagger}$
and $a_i$ are the creation and annihilation operators for the normal
mode $i$ with frequency $\omega_i$, and $g_{nmi}$ are the coupling
parameters of the electron-phonon interaction which we take to be
linear in the phonon normal mode displacement coordinate.

We can separate $H$ into a part that is diagonal with respect to the
electronic degrees of freedom,
\begin{eqnarray} H_0=\sum_n\epsilon_n |n\ket\bra n| +\sum_{ni}g_{nni}|n\ket\bra
n|(a^{\dagger}_i+a_i) +\sum_i\om_ia^{\dagger}_ia_i, \label{H0}
\end{eqnarray} 
and an off-diagonal part $V$ 
\begin{eqnarray} V={\sum_{nmi}}'g_{nmi}|n\ket\bra
m|(a^{\dagger}_i+a_i), \label{V} 
\end{eqnarray}
 where the prime at the summation sign indicates that the terms with
$n=m$ are excluded.  This separation is useful for the following two
reasons. First, in many systems only off-diagonal coefficients
$g_{nmi}$ are small compared to $g_{nni}$.
Hence, $V$ can be treated as a perturbation. Second, for many cases of
interest, the initial density matrix commutes with $H_0$. In this
case, the separation gives simpler forms of the master equations.

For further analysis it is convenient to diagonalize $H_0$ with
respect to the normal mode degrees of freedom. This is achieved with
the unitary operator
\begin{eqnarray}
U=e^{-\sum_{ni}\!\!\frac{g_{nni}}{\om_i}|n\ket\bra
n|(a^{\dagger}_i-a_i)}= \sum_{n}|n\ket\bra
n|e^{-\sum_{ni}\!\!\frac{g_{nni}}{\om_i}(a^{\dagger}_i-a_i)}
\label{unitary} 
\end{eqnarray} 
as 
\begin{eqnarray} \tilde H_0=U^{-1}H_0U
=\sum_n\tilde\epsilon_n |n\ket\bra n|+\sum_i\om_ia^{\dagger}_ia_i,
 \end{eqnarray}
where the renormalized electronic energies are 
\begin{eqnarray}
\tilde\epsilon_n=\epsilon_n-\sum_{i}\frac{g_{nni}^2}{\omega_i}.  
\end{eqnarray}
Applying the same unitary transformation to $V$ gives 
\begin{eqnarray}\tilde
V&=&U^{-1}VU \non \\ &=&{\sum_{nmi}}'\left(g_{nmi}|n\ket\bra m|
\left(a^{\dagger}_i+a_i-\frac{2g_{nni}}{\omega_i}\right)\right. \label{vtilde}
\\ & &\left.\times
e^{\sum_{j}\frac{(g_{nnj}-g_{mmj})}{\om_j}(a^{\dagger}_j-a_j)}\right). \non
\end{eqnarray} Introducing operators \begin{eqnarray} M_{nmi}=g_{nmi}\left(a^{\dagger}_i+
a_i-\frac{2g_{nni}}{\omega_i}\right)e^{\sum_{j}\frac{(g_{nnj}-g_{mmj})}{\om_j}(a^{\dagger}_j-a_j)}
\label{opm}, 
\end{eqnarray}
we can rewrite $\tilde V$ in a compact form as 
\begin{eqnarray}
\tilde V={\sum_{nmi}}'|n\ket\bra m|M_{nmi}.
\end{eqnarray}
The similar unitary transformations were applied to the electron-phonon Hamiltonians 
before \cite{rice, silbey}.
As can be seen from Eq. \ref{vtilde}, in the transformed picture the
electronic transitions from state $|n\ket$ to $|m\ket$ are accompanied
not only by the creation or annihilation of a single phonon of mode
$i$ but also by the displacements of all the normal modes.

Let us take for the initial density matrix 
\begin{eqnarray}
\rho(0)=|n\ket\bra n|
\frac{e^{-\beta\big({\sum_i\om_ia^{\dagger}_ia_i+\sum_{i}g_{nni}(a^{\dagger}_i+a_i)\big)}}}
{{\rm Tr}(e^{-\beta\big({\sum_i\om_ia^{\dagger}_ia_i+\sum_{i}g_{nni}(a^{\dagger}_i+a_i)\big)}})}, \label{initial}
\end{eqnarray}
where the electronic subsystem is prepared in a specific state $|n\ket$ and the
oscillators are in the displaced equilibrium configuration corresponding to this state.
The transformed initial state $\tilde \rho(0)=U^{-1} \rho U$ has the form
\begin{eqnarray}
\tilde\rho(0)=|n\ket\bra n|
\frac{e^{-\beta{\sum_i\om_ia^{\dagger}_ia_i}}}
{{\rm Tr}(e^{-\beta{\sum_i\om_ia^{\dagger}_ia_i}})}. \label{transrho}
\end{eqnarray}
The electronic density matrix in the Eq. \ref{initial}
commutes with unitary operator (\ref{unitary}), hence
its time evolution will be the same in both the transformed and the
untransformed pictures.  For the purpose of developing a perturbative
theory, it is more convenient at this point to work in with the
transformed operators.

\section{Master equation}
We now use the projection operator formalism to obtain the master
equation for the diagonal part of the electronic density matrix
$\rho^{el}$ \cite{Breuer, May}.  For a given projection operator
${\cal P}$ and the initial total density matrix that satisfies
$\rho(0)={\cal P}\rho(0)$, ${\cal P}\rho(t)$ can be shown to satisfy
at least two different master equations: the Nakajima-Zwanzig (NZ) equation
\cite{Nakajima, Zwanzig, Prigogine} and the time-convolutionless (CL)
master equation \cite{Shibata}.
\begin{eqnarray}
\frac{\partial {\cal P}\rho}{\partial t}&=&
-\int_0^td\tau {\cal K}^{NZ}(t-\tau) {\cal P}\rho(\tau), \label{NZ}\\
\frac{\partial {\cal P}\rho}{\partial t}&=&
-\int_0^td\tau{\cal K}^{CL}(\tau) {\cal P}\rho(t)\label{CL}.
\end{eqnarray}
The explicit expressions for superoperators ${\cal K}^{NZ}(\tau)$ and ${\cal K}^{CL}(\tau)$ can 
be found in Ref. \cite{Breuer}. 


Even though  Eqs. \ref{NZ} and \ref{CL} are formally exact,  it is 
impossible to calculate ${\cal K}^{NZ}(\tau)$ and ${\cal K}^{CL}(\tau)$ for most realistic systems.
We will use the perturbation expansion of these superoperators.
To the second order in the coupling constants 
\begin{eqnarray}
{\cal K}_2^{NZ}(\tau)={\cal K}_2^{CL}(\tau)=
{\cal PL}_Ve^{-i{\cal L}_0\tau}{\cal L}_V{\cal P}. \label{comparison}
\end{eqnarray}
Here ${\cal L}_0$ and ${\cal L}_V$ are the Liouville superoperators corresponding
to $\tilde H_0$ and $\tilde V$ whose action on some 
 density matrix $\rho$ is given by
\begin{eqnarray}
{\cal L}_0\rho=\tilde H_0\rho-\rho\tilde H_0, \qquad 
{\cal L}_V\rho=\tilde V\rho-\rho\tilde V.
\end{eqnarray}
In Ref.~\cite{Breuer}, Breuer and Petruccione show 
that to second order in the
coupling constants, the convolutionless experession (Eq. \ref{CL})
gives a better approximation to the exact solution than the
Nakajima-Zwanzig equation (Eq. \ref{NZ}).  It also has an additional
mathematical convenience of being local in time. Therefore, in the
following analysis we will use the convolutionless approach.

Since our initial state is given by Eq. \ref{transrho} we will use the
projection operator that acts on the total density matrix in the
following way
\begin{eqnarray}
{\cal P}\rho=\sum_n|n\ket\bra n|\rho_{eq}^{os}\,
{\rm Tr}\left(|n\ket\bra n|\rho\right), \label{P}
\end{eqnarray}
where 
\begin{eqnarray}
\rho_{eq}^{os}=\frac{e^{-\beta{\sum_i\om_ia^{\dagger}_ia_i}}} {{\rm
Tr}(e^{-\beta{\sum_i\om_ia^{\dagger}_ia_i}})}. \label{equli}
\end{eqnarray}
Using the definition of ${\cal K}_2^{CL}(\tau)$ the following
convolutionless equation is obtained for $P_n=\bra n|\rho_{el}|
n\ket$,
\begin{eqnarray}
\frac{d P_n}{d t}=\sum_mW_{nm}(t)P_m-\sum_mW_{mn}(t)P_{n} \label{Pauli}.
\end{eqnarray}
The coefficients $W_{nm}(t)$ that are, in general, time dependent rates  given by
\begin{eqnarray}
W_{mn}(t)=2\Re e\int_0^td\tau \sum_{ij}\langle M_{nmi}M_{mnj}(\tau)\rangle
e^{-i(\tilde\epsilon_n-\tilde\epsilon_m)\tau}, \label{rates}
\end{eqnarray}
where 
\begin{eqnarray}
\langle M_{nmi}M_{mnj}(\tau)\rangle=
{\rm Tr}\left(M_{nmi}M_{mnj}(\tau)\rho^{os}_{eq}\right) \label{corrfun}
\end{eqnarray} 
and
\begin{eqnarray}
M_{mnj}(\tau)=e^{i\tilde H_0\tau}M_{mnj}e^{-i\tilde H_0\tau}.
\end{eqnarray} 
Due to the explicit form of operators $M_{nmi}$ (Eq. \ref{opm}) the
calculation of the correlation functions in Eq. \ref{corrfun} can be
reduced to the averaging of the displacement operators over the
equilibrium ensemble (Eq. \ref{equli}). After straightforward, but
lengthy, calculations we obtain the principal result of this paper:
\begin{eqnarray}
& &\langle M_{nmi}M_{mnj}(\tau)\rangle=g_{nmi}g_{mnj} \non \\
& &\times\left(\left(\Delta_{nmi}(\overline{n}_i+1)e^{i\om_i\tau}
-\Delta_{nmi}\overline{n}_ie^{-i\om_i\tau}+\Omega_{nmi}\right) \right.\non \\
& &\times\left.\left(\Delta_{nmj}(\overline{n}_j+1)e^{i\om_j\tau}
-\Delta_{nmj}\overline{n}_je^{-i\om_j\tau}+\Omega_{nmj}\right)\right.\non \\
& &\left.+\delta_{ij}(\overline{n}_i+1)e^{i\omega_i\tau}+\delta_{ij}\overline{n}_i
e^{-i\omega_i\tau}\right)q_{nm}(\tau)f_{nm}(\tau).
 \label{corrf}
\end{eqnarray}
Here 
\begin{eqnarray}
\Delta_{nmi}&=&\frac{(g_{nni}-g_{mmi})}{\om_i}, \\
\Omega_{nmi}&=&\frac{(g_{nni}+g_{mmi})}{\om_i}, \\
\overline{n}_i&=&\frac{1}{e^{\beta\om_i}-1}, \\
q_{nm}(\tau)&=&e^{i\sum_j{\Delta^2_{nmj}\sin\omega_j\tau}},\label{q} \\ 
f_{nm}(\tau)&=&e^{-2\sum_j(\overline{n}_j+\frac{1}{2})\Delta_{nmj}^2(1-\cos\om_j\tau)}. \label{f}
\end{eqnarray}
In the case when all the diagonal electron/phonon terms vanish,
 $g_{nni}=0$, the correlation functions in Eq. \ref{corrf} reduce to
 those obtained within the golden rule approach \cite{May}. It is
 clear from Eq. \ref{corrf} that for systems with $g_{nni}/\omega_i
 \gg 1$
 the golden rule approach is not applicable.

The qualitative behavior of the probabilities $P_n(t)$ depends on the
explicit form of the time dependent rates $ W_{mn}(t)$ in
Eq. \ref{Pauli}. Depending on a particular system, certain
simplifications of $W_{mn}(t)$ are possible.  In particular, for large
systems we can expect the integrand in Eq. \ref{rates} to decay to
zero and, correspondingly, $W_{mn}(t)$ to become constant in the long
time limit. If the typical electronic relaxation times are much longer
then the decay rate of the integrand in Eq. \ref{rates}, then the
$W_{mn}(t)$ can be replaced by their time independent Markovian
limits. Due to the rather complicated functional forms of $q_{nm}(t)$
and $f_{nm}(t)$ in Eq. \ref{corrf} we could not obtain an explicit
expression for $W_{mn}(t)$ in the Markovian limit. However, for models
that satisfy the condition
$\sum_j(\overline{n}_j+\frac{1}{2})\Delta_{nmj}^2\gg 0$ and with
$\Delta_{nmi}$ taken as smooth function of $i$, we can approximate
$f_{nm}(\tau)$ in Eq. \ref{f} by a Gaussian, viz.,
\begin{eqnarray}
f_{nm}(\tau)\approx e^{-a_{nm}\tau^2} 
\end{eqnarray}
where
\begin{eqnarray} 
a_{nm}=\sum_j(\overline{n}_j+\frac{1}{2})\Delta_{nmj}^2\om_j^2. 
\end{eqnarray}
For $\om_i\tau \ll 1$, we can also approximate the $\sin \om_i \tau$
terms that appear in functions $q_{nm}(t)$ (Eq. \ref{q}) by $\sin
\om_i \tau \approx \om_i \tau$ so that
\begin{eqnarray}
q_{nm}(t)\approx e^{i\gamma_{nm}t},
\end{eqnarray} 
where 
\begin{eqnarray} 
\gamma_{nm}=\sum_i\Delta_{nm}^2\omega_i.
\end{eqnarray}
Under these approximations the long time (Markovian) limit of
$W_{nm}(t)$ can be obtained as
\begin{eqnarray}
\lim_{t\to\infty}W_{nm}(t)=\sum_{ij}L_{nmij},
\end{eqnarray}
where
\begin{widetext}
\begin{eqnarray}
L_{nmij}&=&g_{nmi}g_{mnj}
\biggl(\Delta_{nmi}\Delta_{nmj}\Bigl((\overline{n}_i+1)(\overline{n}_j+1)
G_{nm}(\om_i+\om_j+\gamma_{nm}-\tilde\epsilon_n+\tilde\epsilon_m) \non  \\ 
& &+\overline{n}_i\overline{n}_j
G_{nm}(-\om_i-\om_j+\gamma_{nm}-\tilde\epsilon_n+\tilde\epsilon_m)
-(\overline{n}_i+1)\overline{n}_j
G_{nm}(\om_i-\om_j+\gamma_{nm}-\tilde\epsilon_n+\tilde\epsilon_m)\non \\
& &-(\overline{n}_j+1)\overline{n}_i
G_{nm}(\om_j-\om_i+\gamma_{nm}-\tilde\epsilon_n+\tilde\epsilon_m)\Bigr)\non \\
& &+\Delta_{nmi}\Omega_{nmj}\Bigl((\overline{n}_i+1)
G_{nm}(\om_i+\gamma_{nm}-\tilde\epsilon_n+\tilde\epsilon_m)- \overline{n}_i
G_{nm}(-\om_i+\gamma_{nm}-\tilde\epsilon_n+\tilde\epsilon_m)\Bigr) \non \\
& &+\Delta_{nmj}\Omega_{nmi}\Bigl((\overline{n}_j+1)
G_{nm}(\om_j+\gamma_{nm}-\tilde\epsilon_n+\tilde\epsilon_m)- \overline{n}_j
G_{nm}(-\om_j+\gamma_{nm}-\tilde\epsilon_n+\tilde\epsilon_m)\Bigr) \non \\
& &+\Omega_{nmi}\Omega_{nmj}G_{nm}(\gamma_{nm}-\tilde\epsilon_n+\tilde\epsilon_m)\non \\
& &+\delta_{ij}(\overline{n}_i+1)G_{nm}((\om_i+\gamma_{nm}-\tilde\epsilon_n+\tilde\epsilon_m) \non \\
& &+\delta_{ij}\overline{n}_iG_{nm}(-\om_i+\gamma_{nm}-\tilde\epsilon_n+\tilde\epsilon_m) \biggr),
\end{eqnarray}
\end{widetext}
with 
\begin{eqnarray}
G_{nm}(x)=\sqrt{\frac{\pi}{a_{nm}}}\exp{\left(-\frac{x^2}{4a_{nm}}\right)}.
\end{eqnarray}

\section{Relaxation in model electron-phonon systems}

In what follows, we examine the solution of the Pauli equation for a
number of model systems using the correlation functions (Eq.
\ref{corrf}) defined above.  We take two examples, one in consisting
of a two state electronic system coupled to a single acoustic phonon
branch with Ohmic and super-Ohmic couplings.  The other taken from our
recent work in describing secondary exciton formation at the domain
boundary in organic polymer semiconductor heterojunctions.  We also,
then compare the results obtain using the present method to those
obtained using Marcus' formulation of the rate-constant.

\subsection{Ohmic vs. Super-Ohmic Coupling}

As our first example we consider a system with two electronic levels
coupled to $N$ normal modes with frequencies $\omega_i =
A\sin(i\pi/2N)$ with $i\in[ 1,N]$ that model the acoustic phonon
branch.
 We will choose the coupling coefficients in the following form
$g_{12i}=g_{21i}=g_{11i}=\lambda (\omega_i)^p$ and $g_{22i}=-\lambda
(\omega_i)^p$, where $\la$ and $p$ are parameters.  We will consider
the cases of $p=-1/2$ and $p=0$ that correspond approximately to the
Ohmic and super-Ohmic baths in the golden rule approach to relaxation.

\begin{figure}
 \includegraphics[width=\columnwidth]{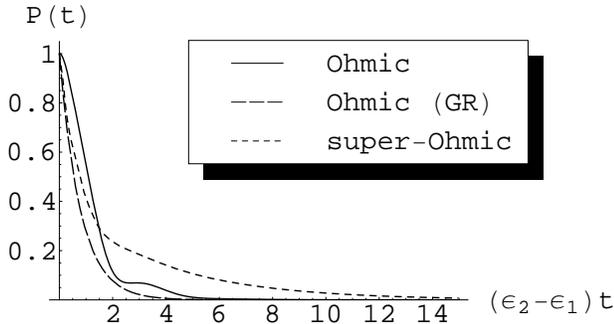}  
 \caption{\label{Figure1} The population of the excited state as a
 function of time for the $80$ mode system with
 $\omega_j=A\sin(\frac{j\pi}{160})$, The solid line corresponds to
 $A=10$, $\la=0.1$, $p=-1/2$, and $\epsilon_2-\epsilon_1=1$, the line
 of short dashes corresponds to $A=10$, $\la=0.2$, $p=0$, and
 $\epsilon_2-\epsilon_1=1$. The line of long dashes corresponds to the
 golden rule results for the Ohmic case. Temperature is $298$ K for
 all cases.}
 \end{figure}

\begin{figure}
 \includegraphics[width=\columnwidth]{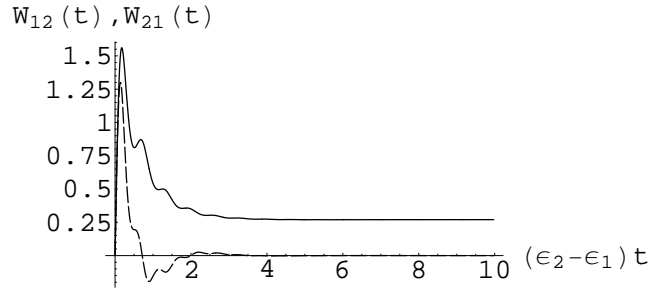}  
 \caption{\label{Figure2} Transition rates $W_{12}(t)$ and $W_{21}(t)$ for the super-Ohmic model} 
 \end{figure}

Taking the initial state to be prepared in the higher-lying electronic
state, electronic population of the higher electronic state as a
function of time is shown on Fig. \ref{Figure1}. It can be seen that
the relaxation does not follow a simple exponential decay.  This
behavior can be understood by looking at the explicit form of the rate
coefficients $W_{12}(t)$ and $W_{21}(t)$ as functions of time as shown
on Fig. \ref{Figure2} for the super-Ohmic case.  Only after the rate
coefficients become nearly constant in time does exponential
relaxation occur. For the chosen parameter values, substantial
non-exponential relaxation occurs before that time.  Note the for the
chosen parameters the complete electronic relaxation occurs before the
onset of any possible recurrence phenomena due to the finite number of
modes.  Such recurrences will occur at the time scale of the order of
$2\pi$ divided by the smallest normal mode frequency. It can also be
seen on Fig. 2 that coefficients $W_{12}(t)$ and $W_{21}(t)$ can
acquire negative values. This is in contrast to the golden rule
approach where the coefficients are always positive and time
independent. Because of the time dependence of coefficients
$W_{12}(t)$ and $W_{21}(t)$ and their approximate nature the
time-convolutionless master equation is not guaranteed to preserve the
positivity of probabilities.  Fig. 2 also shows the electronic
relaxation for the Ohmic model obtained by applying the golden rule to
the Hamiltonian (\ref{Ham}) when coefficients $g_{nni}$ are treated on
the same basis as $g_{nmi}$ ($n\neq m$). No reliable golden rule
results can be obtained for the super-Ohmic case because of the
relatively small number of modes in the model.

\subsection{Exciton regeneration at a semiconductor interface}

As another example we consider the electronic relaxation in the
conjugated polymer heterojunctions previously investigated in
Ref. \cite{Ramon1, Ramon3, Ramon2} by a different approach.  
Here we consider the decay of an excitonic state into a charge-separated
state at the interface between two semiconducting polymer phases (TFB) and  
(F8BT). Such materials have been extensively studied for their potential in 
organic light-emitting diodes and organic photovoltaics
\cite{morteani:247402,russell:2204,silva:125211}.
 At the phase boundary, the material forms a type-II semiconductor
heterojunction with the off-set between the valence bands of the two
materials being only slightly more  than the binding energy of 
an exciton placed on either the TFB or F8BT polymer.  As a result, an exciton on 
the F8BT side will dissociate to form a charge-separated (exciplex) state at the 
interface. 
$$
{\rm F8BT}^*:{\rm TFB}\longrightarrow {\rm F8BT}^-:{\rm TFB}^+.
$$
Ordinarily, such type II systems are best suited for photovoltaic rather than LED 
applications
However, LEDs fabricated from phase-segregated 50:50 blends of TFB:F8BT 
give remarkably efficient electroluminescence efficiency due to {\em secondary} 
exciton formation due the back-reaction
$$
{\rm F8BT}^-:{\rm TFB}^+\longrightarrow {\rm F8BT}^*:{\rm TFB}.
$$ 
as thermal equilbrium between the excitonic and charge-transfer
states is established.  This is evidenced by long-time emission,
blue-shifted relative to the emission from the exciplex, accounting
for nearly 90\% of the integrated photo-emission.

Here, as above, as consider only two lowest electronic levels corresponding to 
$$
|XT\rangle =   {\rm F8BT}^*:{\rm TFB} \,\,\&\,\, |CT\rangle= {\rm F8BT}^-:{\rm TFB}^+.
$$ 
As was shown in Ref. \cite{Bittner1, Bittner2, Bittner3} for such
systems there are two groups of phonon modes that are coupled strongly
to the electronic degrees of freedom as evidenced by their presence as
vibronic features in the vibronic emission spectra, namely: low
frequency torsional modes along the backbones of the polymer and
higher frequency C=C stretching modes.  The normal modes within each
group have about the same frequencies and form two separated 
bands.  The interested reader is referred to Ref. \cite{Bittner1,
Bittner2, Bittner3} for specific details regarding the
parameterization of the model.

\begin{figure}
 \includegraphics[width=\columnwidth]{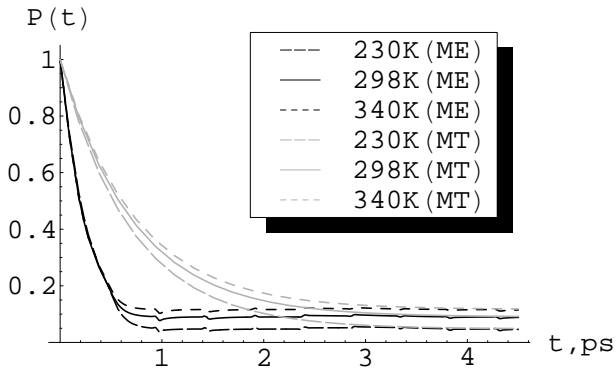}  
 \caption{\label{Figure3}The population of the excited state as a
 function of time for the TFB:F8BT heterojunction for three different
 temperatures obtained by solving the time-convolutionless master
 equation (ME) or through the Marcus-type approximation (MT).}
 \end{figure}

\begin{figure}
 \includegraphics[width=\columnwidth]{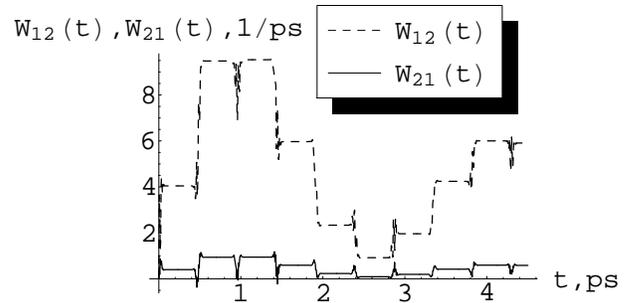}  
 \caption{\label{Figure4}Transition rates $W_{12}(t)$ and $W_{21}(t)$ for the TFB:F8BT heterojunction (see
 Ref. \cite{Ramon1} for the model details) involving  24 normal modes  at $298$ K.} 
 \end{figure}

As before, we consider the initial state as being prepared in the
$|XT\rangle$ state corresponding to photoexcitation of the F8BT
polymer.  The results for the higher electronic level population
obtained by numerically solving Eq. (\ref{Pauli}) are shown on
Fig. \ref{Figure3}.  The time dependence of coefficients $W_{12}(t)$
and $W_{21}(t)$ is shown on Fig. \ref{Figure4}.  As in the previous
example the electronic relaxation does not follow a simple exponential
pattern.  The main difference between this model and the previous
example is that the electronic relaxation time is of the same order as
the recurrence time for coefficients $W_{12}(t)$ and $W_{21}(t)$. As
can be seen from Fig. \ref{Figure4} the time dependence of these
coefficients has the form of approximately constant regions abruptly
changed at regular recurrence intervals. This type of time dependence
makes the electronic relaxation look like a series of exponential
relaxations with changing rates. As in the previous example,
coefficients $W_{12}(t)$ and $W_{21}(t)$ can become negative as the 
relaxation proceeds towards the equilibrium population. 

Note also, that the because the relaxation does not obey a simple 
exponential rate law, the initial decay slightly ``overshoots'' the final 
equilibrium population.  This is most evident in the highest temperature 
case considered here (T = 340K).   Since the $|XT\>$ is also  the emissive 
state, photo-emission (not included herein) depletes this population on a
nanosecond time scale (radiative lifetime).  
Population  back-transfer from the $|CT\>$ to maintain a thermal equilibrium population
then leads to the continuous replenishment of the emissive species
such that nearly all of the CT population contributes to the formation of 
secondary (regenerated) excitonic states
\cite{morteani:247402,russell:2204,silva:125211,Ramon1,Ramon2,Ramon3}.

It is of interest to compare the relaxation dynamics in TFB:F8BT
heterojunction obtained through the application of the
time-convolutionless master equation with other approaches. As an
example we will consider the Marcus-type approach which is widely used
to study electron transfer in chemical systems \cite{May, Ramon3,
Nitzan, Stuch}.  
Because of the coordinate dependency of the coupling, some care must be 
taken in deriving the Marcus rates and we give details of this in the 
appendix.
The results for the relaxation dynamics using these
rates applied to TFB:F8BT heterojunction for three different
temperatures are shown in Fig. \ref{Figure3}.  It can be seen that
apart from a more complicated time dependence the master equation
approach gives faster relaxation when compared to the Marcus-type
picture. The discrepancy between the two approaches can be explained
by the fact that the Marcus approximation is assumed to be valid when
$kT \gg \hbar \omega_i$ for all the normal modes. In the case of the
TFB:F8BT heterojunction this condition is not satisfied for the higher
frequency modes.

\section{Conclusions}
We derived the explicit form of the time-convolutionless master
equation for the electronic populations in the electron-phonon
systems. We applied it to study the electronic relaxation in the
two-level electronic systems coupled to the Ohmic and super-Ohmic
baths, as well as in the realistic conjugated polymer
heterojunction. The time evolution of the electronic populations is,
in general, more complex then the one obtained with the Pauli master
equation and depends on the explicit form of the time dependent
coefficient. The time-convolutionless master equation can account for
the appearance of recurrence effects in smaller electron-phonon
systems.

\begin{acknowledgments}
This work was funded in part through grants from the National Science
Foundation and the Robert A. Welch foundation.  Seed funds from the 
Texas Center for Superconductivity (TcSUH) are also acknowledged. 
\end{acknowledgments}

\appendix*
\section{Derivation of Marcus-type rate expressions}
To apply the Marcus approach for the system at hand, it is convenient to rewrite
Hamiltonians (\ref{H0}) and (\ref{V}) in terms of the coordinate $q_i$
and momentum $p_i$ operators for the normal modes as
\begin{eqnarray}
H_0=\sum_n\epsilon_n |n\ket\bra n|
+\sum_{ni}\tilde g_{nni}|n\ket\bra n|q_i +\sum_i\frac{p_i^2}{2}+\sum_i\frac{\om_i^2q_i^2}{2},  
\end{eqnarray}
and an off-diagonal part $V$
\begin{eqnarray}
V={\sum_{nmi}}'\tilde g_{nmi}|n\ket\bra m|q_i, 
\end{eqnarray}
where 
\begin{eqnarray}
\tilde g_{nmi}=\sqrt{2\omega_i}g_{nmi}.
\end{eqnarray}
The coordinate and momentum operators are then treated as static
parameters and the rate constants for the transition from state
$|n\ket$ to state $|m\ket$ is given by averaging the golden rule
transition rates over the initial equilibrium distribution of the
coordinates corresponding to the equilibrium ensemble for Hamiltonian
$H_0$ assuming the that the electronic state of the system is $|n\ket$
\cite{May}, i.e.,
\begin{widetext}
\begin{eqnarray}
k_{nm}=2\pi \int d\{q_i\}f(\{q_i\})|V_{nm}(\{q_i\})|^2
\delta\Big(U_n(\{q_i\})-U_m(\{q_i\})\Big), \label{rate}
\end{eqnarray}
\end{widetext}
where $\{q_i\}$ denotes all coordinate variables and 
\begin{eqnarray}
f(\{q_i\})=\frac{1}{Z}e^{-\beta U_n}
\end{eqnarray}
and 
\begin{eqnarray}
V_{nm}(\{q_i\})=\sum_i\tilde g_{nmi}q_i.
\end{eqnarray}
Here $Z$ is the partition function and $U_n$ is given by 
\begin{eqnarray}
\epsilon_n 
+\sum_{i}\tilde g_{nni}q_i +\sum_i\frac{\om_i^2q_i^2}{2}.
\end{eqnarray}
Note that Eq. \ref{rate} differs from the similar expression of
Ref. \cite{May} by the presence of coordinate dependence in the
interaction parameters $V_{nm}(\{q_i\})$. The integrations in
Eq. \ref{rate} can be explicitly performed giving
\begin{eqnarray}
k_{nm}&=&\left(\left(\frac{\Delta E_{nm}P_{nm}}{E_{nm}^\lambda} +F_{nm}\right)^2
-\frac{2G_{nm}^2}{\beta E_{nm}^\lambda}+\frac{H_{nm}}{\beta}\right) \non \\
& &\times \sqrt{\frac{\pi \beta}{E_{nm}^\lambda}}
\exp\left(-\frac{\beta(\Delta E_{nm}-E_{nm}^\lambda)^2}{4E_{nm}^\lambda}\right). \label{exrate}
\end{eqnarray}
Here $\Delta E_{nm}$ and $E_{nm}^\lambda$ 
are given by 
\begin{eqnarray}
\Delta E_{nm}&=&\epsilon _n-\epsilon_m-\sum_i\frac{\tilde g_{nni}^2}{2\omega_i^2}
+\sum_i\frac{\tilde g_{mmi}^2}{2\omega_i^2},  \\
E_{nm}^\lambda&=&\sum_i\frac{(\tilde g_{nni}-\tilde g_{mmi})^2}{2\omega_i^2}
\end{eqnarray}
and are usually referred to as the driving force and reorganization
energy, respectively.  Parameters $P_{nm}$, $F_{nm}$, and $H_{nm}$ are
given by
\begin{eqnarray}
P_{nm}&=&\frac{\tilde g_{nmi}(\tilde g_{nni}-\tilde g_{mmi})}{2\omega_i^2}, \\
F_{nm}&=&\frac{\tilde g_{nmi}(\tilde g_{nni}+\tilde g_{mmi})}{2\omega_i^2},\\
H_{nm}&=&\frac{\tilde g_{nmi}^2}{\omega_i^2}.
\end{eqnarray}
It can be verified from Eqs. (\ref{rate}) or (\ref{exrate}) that  rates $k_{nm}$ 
satisfy the principle of the detailed balance, i. e.,
\begin{eqnarray}
\frac{k_{nm}}{k_{mn}}=e^{\beta\Delta E_{nm}}.
\end{eqnarray}

\bibliography{Master}

\end{document}